# Probing Emergent Surface and Interfacial Properties in Complex Oxides via in situ X-ray Photoelectron Spectroscopy

Suresh Thapa[+], Rajendra Paudel[+], Miles D. Blanchet[+], Patrick T. Gemperline[+], and Ryan B. Comes*

Department of Physics, Auburn University, Auburn, AL 36849

+Equal contributions

*Corresponding Author, E-mail: ryan.comes@auburn.edu

## Abstract

Emergent behavior at complex oxide interfaces has driven much of the research in the oxide thin film community for the past twenty years. Interfaces have been engineered for potential applications in spintronics, topological quantum computing, and high-speed electronics in cases where the bulk materials would not exhibit the desired properties. Advances in thin film growth have made the synthesis of these interfaces possible, while surface characterization tools such as X-ray photoelectron spectroscopy have been critical to understanding surface and interfacial phenomena in these materials. In this review we discuss the leading research in the oxide field over the past 5-10 years with a focus on connecting the key results to the X-ray photoelectron spectroscopy studies that enabled them. We describe how *in situ* integration of synthesis and spectroscopy can be used to improve the film growth process and to perform immediate experiments on specifically tailored interfacial heterostructures. These studies can include determination of interfacial intermixing, valence band alignment, and interfacial charge transfer. We also show how advances in synchrotron-based spectroscopy techniques have answered questions that cannot be addressed in a lab-based system. By further tying together synthesis and spectroscopy through *in situ* techniques, we conclude by discussing future opportunities in the field through the careful design of thin film heterostructures that are optimized for X-ray studies.

## I. Introduction

From multiferroics to two-dimensional electron gases (2DEGs) to strongly correlated and topological systems, much of complex oxide thin film research has been driven by interfacial phenomena for the past 20 years. Groups have demonstrated emergent oxide 2DEGs due to polar/non-polar interfaces[1], multiferroic behavior with magnetoelectric control from interfaces between oxide ferroelectrics and ferromagnets[2,3], and emergent ferromagnetism[4] and orbital polarization[5] due to interfacial charge transfer. Numerous groups have reviewed the novel properties that have been predicted and reported at oxide interfaces, and readers are referred to several recent review articles summarizing the state-of-the-art in oxide thin film research[6–8]. Advances in thin film synthesis continue to drive the field as well, including new approaches to molecular beam epitaxy[9–11] and pulsed laser deposition[12,13] to enable more precise control of cation stoichiometry than has been possible previously. Such approaches are highly complementary with materials characterization techniques that help to explain the physical origins of emergent phenomena in oxide thin films and interfaces.

In this review, we will discuss how the reported interfacial phenomena can be examined using surface science tools—particularly *in situ* X-ray photoelectron spectroscopy (XPS)—in concert with film synthesis and other *ex situ* techniques to best understand the properties of oxide thin films, surfaces, and interfaces, as shown conceptually in Figure 1. Beginning with several recent examples from the literature that motivate the importance of *in situ* characterization, we then present a brief overview of the physics of XPS and how to design experiments that best integrate thin film synthesis and *in situ* XPS to achieve high impact scientific results. We then present several examples of how XPS has been used in concert with film synthesis to better understand surface properties, interfacial charge transfer, electronic band alignment across a heterojunction, and non-idealities such as intermixing and off-stoichiometry. Finally, we present an overview of future opportunities that can leverage user facilities and the lessons learned from lab-based *in situ* studies for further breakthroughs.

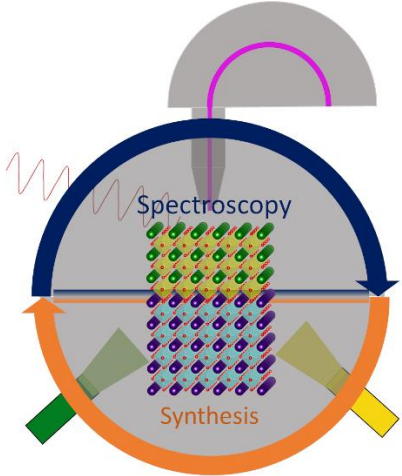

*Figure 1: Feedback loop of in situ synthesis and X-ray photoelectron spectroscopy for study of oxide thin films and interfaces.*

## A. Overview of Complex Oxide Thin Films and Interfaces

The pursuit of a high-mobility 2DEG at complex oxide interfaces has driven research in a wide variety of materials for more than a decade. Such 2DEGs offer the promise of higher carrier concentrations than can be achieved with other materials systems, making them exciting candidates for high speed electronics. Oxide 2DEGs also offer additional benefits over their semiconductor counterparts in that they can exhibit strong spin-orbit coupling when comprised of 5$d$ electrons, such as those in $KTaO_3$[14]. Spin-orbit coupling in a 2DEG offers an additional opportunity not available in conventional semiconductor devices. Beginning with the $LaAlO_3/SrTiO_3$ interface[1], many candidate materials systems have been explored. The $LaAlO_3/SrTiO_3$ system was proposed for transistors with high mobility and high carrier-concentrations[15,16], spin-orbit coupled heterostructures with Rashba splitting[17,18], and numerous other applications. However, controversy regarding the role of defects in the $LaAlO_3/SrTiO_3$ system[19–21] has been present for many years. As we will discuss later, XPS along with a variety of other complementary characterization techniques helped to explain the physical origin of the 2DEG in these materials[19,22,23].

Subsequent materials research has focused on alternative structures that do not involve polar/non-polar interfaces, including interfaces between STO and rare earth titanates, such as in $NdTiO_3/SrTiO_3$,[24,25] $GdTiO_3/SrTiO_3$,[26] and $LaTiO_3/SrTiO_3$.[27,28] Unlike in polar/non-polar interfaces, here delta-doping of electrons produces the 2DEG, with the rare earth A-site ion providing an additional electron for the system. Emergent phenomena have also been reported in these interfacial materials, including superconductivity[28] and ferromagnetism.[29,30] As with the $LaAlO_3/SrTiO_3$ system, however, defects can still play a significant role, albeit in a different fashion. Various groups have reported excess oxygen content in rare earth titanates[25,31,32] due to either imperfect tuning of oxygen composition during growth or atmospheric exposure after growth, which has been confirmed through XPS in each case. These changes in oxygen stoichiometry can have profound impacts on material properties.

With the emerging emphasis on materials for quantum computation, materials that exhibit strong spin-orbit coupling have generated an increased interest in the oxide community as well. Work has focused on 5d transition metal oxides materials such as KTaO$_3$[14,33], which exhibits the Rashba effect, and SrIrO$_3$, which possesses stronger spin-orbit coupling. In the case of KTaO$_3$, surface 2DEGs have been reported in ultra-high vacuum conditions from oxygen vacancies[33,34]. Since these reports, a variety of means have been developed to produce interfacial 2DEGs in KTaO$_3$ primarily through the introduction of polar/non-polar interfaces[35–37]. These interfaces have exhibited such interesting properties as an optically-tunable Rashba effect and greater Hall mobility than analogous LaAlO$_3$/SrTiO$_3$ interfaces due to the wider bandwidth of 5$d$ electrons than 3$d$ electrons[36]. As in previous examples, however, a more complete understanding of KTaO$_3$ properties was gleaned from an *in situ* cleaved single crystal through a combination of scanning tunneling microscopy (STM), ion scattering spectroscopy (ISS), and XPS[38]. In this work, the authors showed that to compensate the polar KTaO$_3$ surface, a series of physical and chemical distortions occur and that the most stable surface is passivated by water adsorption.

Perovskite iridates are another class of 5d oxides that have generated excitement for their potential in materials systems taking advantage of topological phenomena[39,40]. SrIrO$_3$ is a semimetallic oxide that exhibits strong spin-orbit coupling with evidence of a dimensionality driven cross-over between metallic and semiconducting behavior in ultrathin films[41,42]. In oxide heterostructures, SrIrO$_3$ interfaces have been studied extensively for magnetic and electronic phenomena that occur due to charge transfer and electron correlations. In STO/SrIrO$_3$ superlattices, others have shown the emergence of an energy gap at the Fermi level as the thickness of the iridate layers is reduced to ~4 u.c.[43] They also observed the formation of electronic energy features resembling a Dirac-cone, suggesting that topological effects occur in the ultra-thin limit. (111)-oriented SrIrO$_3$/SrTiO$_3$ heterostructures have also been synthesized and proposed as topological materials that exhibit the desirable honeycomb lattice for topological quantum information systems[44]. Similar (100)-oriented SrIrO$_3$/SrTiO$_3$ superlattices have also been shown to behave like layered Ruddelsden-Popper iridates by reducing the iridate layer to two unit cells[45]. The SrMnO$_3$/SrIrO$_3$ interface has also been studied through superlattice synthesis and shown to exhibit ferromagnetism due to electron transfer from Ir to Mn across the interface, leading to the anomalous Hall effect in these structures[46,47]. LaMnO$_3$/SrIrO$_3$ interfaces have also demonstrated the Rashba effect, with XPS and X-ray absorption measurements confirming electron transfer from Ir to Mn[48].

Strongly correlated oxides such as nickelates have been a long-standing area of interest within the oxide community, due to their potential to exhibit superconductivity that is analogous to the layered cuprates. Efforts have broadly focused on methods to induce orbital polarization in these materials to create a quasi-two-dimensional plane of nickel ions that resembles the planar cuprate structure. Groups have shown that through by heterostructuring LaNiO$_3$ into a superlattice with LaTiO$_3$, orbital polarization can be induced to break the symmetry of the Ni $e_g$ energy levels[5,49]. We will discuss ways to examine this type of charge transfer by *in situ* XPS later in this article. Nickelate thin films have also been shown to exhibit conductivity that is dependent on film thickness and surface termination[50–54]. Theoretical predictions have suggested that this phenomenon is due to structural distortions in the surface layer[51], but recent work employing XPS and X-ray absorption spectroscopy (XAS) has suggested that surface oxygen vacancies could also be the cause[55]. Recently, an infinite-layer (Nd,Sr)NiO$_2$ nickelate heterostructure was demonstrated to be superconducting for the first time[56–58], though this result has not yet been independently confirmed to our knowledge. Future studies of superconducting nickelates using X-ray spectroscopy will hopefully provide further insights into the very exciting initial results.

In summary, epitaxial complex oxide thin films continue to be a fruitful area of exploration in the condensed matter physics and fundamental materials science communities. However, the complexity of the perovskite oxide materials system in terms of bulk defect tolerance and non-idealities at film surfaces and interfaces makes careful materials characterization vital to best explain the wide range of emergent properties that have been reported in oxide heterostructures. Our goal in this review is to show how XPS measurements have been used in the past to answer many of these questions such that the reader may better be able to employ the versatile technique in future studies. We also aim to provide further insights and interpretations on the current state-of-the-art in the oxide interface literature, which is awash in exciting breakthroughs but also is prone to misinterpretation of XPS and complementary spectroscopy data.

### B. Physics of X-ray Photoelectron Spectroscopy (XPS)

The photoelectric effect is the phenomenon at the core of XPS measurements. It is so called because photons propagating through a material are absorbed by core-level and valence electrons in atoms. The resulting energy of the excited electrons is enough to overcome the electrostatic potential of the material and escape into vacuum. The photoelectric effect was first discovered in 1887 by Heinrich Hertz[59], but it was not until 1905 when Albert Einstein integrated the quantization of the light into the photo electric effect that it was fully understood[60]. Einstein's discovery that the energy is $E_{photon}$ = hv, where h is Plank's constant and v is the frequency of light, subsequently proved the wave particle duality of light. Kai Siegbahn later won the Nobel Prize for his invention of XPS (then referred to as "electron spectroscopy for chemical analysis")[61].

Assuming we precisely control the wavelength of light using a Rowland circle monochromator, we may determine the binding energy of an electron via the equation:

$$E_{photon} = hv = E_B + \Phi_S + E'_K. \quad (1)$$

where the binding energy, $E_B$, is the energy needed to bring it to the Fermi level, the material's work function, $\Phi_S$ is the energy needed to free an electron at the Fermi level and get it to vacuum, and the vacuum kinetic energy is $E'_k$. XPS works by directing an X-ray beam generated by a monochromatic X-ray source onto a sample as show in Figure 1. The photoelectric effect causes bound electrons to be ejected out of the sample where they pass through an analyzer. This analyzer uses electron optics to select only electrons with a certain kinetic energy. By varying the settings of the optics, the XPS can sweep a range of energies. Modern XPS systems employ analyzers that allow for several electron energies to be detected simultaneously, thus making data collection more efficient[62–64]. In order to connect the photon energy and the kinetic energy of the electrons measured by the XPS, it is best to think about electronic band alignment diagram of XPS in Figure 2.

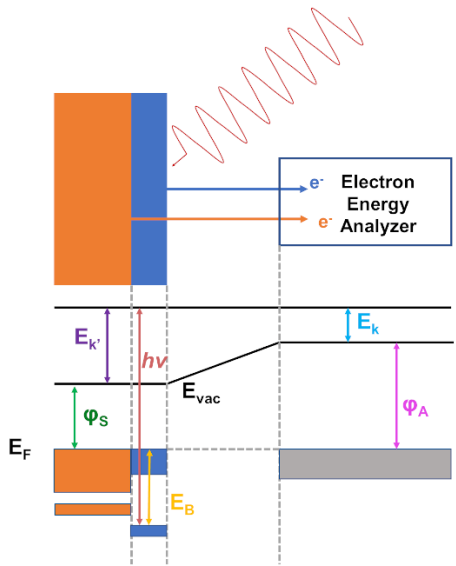

*Figure 2: Energy schematic for photoemission from a thin film heterostructure.*

There are techniques to determine the sample's work function, but there is an immediate problem in that the kinetic energy measured by the detector, $E_K$, is not the same kinetic energy of the emitted electrons, $E_{K'}$. Due to biasing of the optics and electron energy analyzer, there is a work function of each XPS, $\Phi_A$, that relates the $E_K$ to $E_{K'}$. However, if the sample is mounted on a stage that shares a common ground with the XPS analyzer, then we can equate the Fermi levels of the sample and XPS. Now by looking at the band diagram, we can see that $E_{K'} = E_K + \Phi_A - \Phi_S$. Substituting this into equation 1, we get $h\nu = E_B + E_K + \Phi_A$ or $h\nu = E_B + E_K + \Phi_A$. So by using the optics of an XPS to select and count only electrons with $E_K$ and by using controlling the X-ray source frequency, we are able to calculate the binding energy of each electron excited by the photoelectric effect without knowing the sample work function, $\Phi_S$. The probability of photoelectrons escape from the material is also sensitive to the angle of emission from a planar surface and the kinetic energy of the electron. These features are governed by the inelastic mean free path, $\lambda$, of an electron within the material[65], which can be modeled by the equation:

$$\lambda(\text{nm}) = \frac{143}{E_{k'}^2 \text{ (eV}^2)} + 0.054 * \sqrt{E_{k'}(\text{eV})}. \quad (2)$$

While equations (1) and (2) describe the measurement of electron binding energy via XPS, separate physical phenomena govern the observation of peaks at various binding energies. Of these phenomena, spin-orbit coupling or core-level electrons, multiplet splitting, and satellite peaks are some of the most commonly observed when it comes to understanding XPS data, though these are not always accurately interpreted[66]. Each of these will be covered briefly in this section. Auger electrons are also detected during an XPS measurement and can be beneficial for elemental identification but are also a source of frustration if the kinetic energy of the Auger electron is close to that of an elemental core level.

The first effect to consider is that of spin-orbit coupling. Electrons in the same orbital do not all have the same binding energy. This is because spin-up and spin-down electrons have different total angular momenta. The energy shift caused by this spin-orbit coupling goes as $L * S$. Thus for all orbitals where l ≠0, there will be a splitting of the binding energy for spin-up and spin-down electrons. The two peaks correspond to the two possible values of total angular moment of the photoelectrons, $j = l \pm s$. For a 2p electron, the total angular momentum can take either a value of $j=1+½=3/2$ or $j=1−1/2 =1/2$. This phenomenon can be seen in the two largest peaks of Figure 3 for the Co 2p peak. Additionally, the ratio of the intensities of the peaks are seen not to be 1:1 but is instead 2:1. This is because the relative intensities are determined by the number of magnetic sub-state configurations, $2j + 1$, that correspond to each $j$ value. Knowing the expected ratio of the peak intensities is important when fitting XPS spectra as it allows for other effects to be distinguished from the spin-orbit coupling.

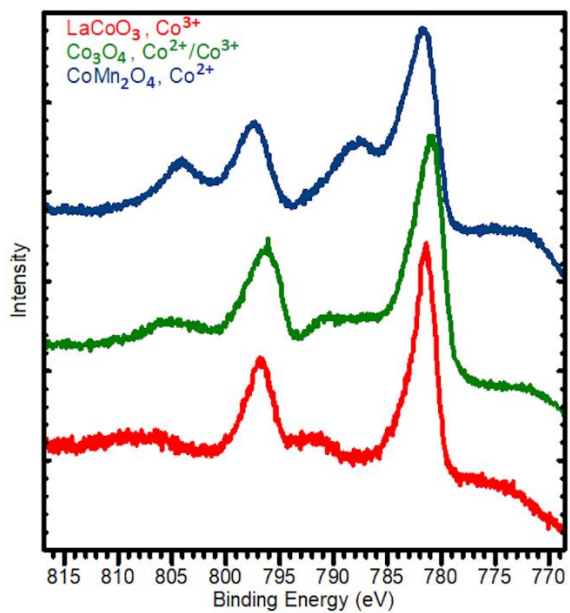

*Figure 3: Co 2p XPS data for LaCoO$_3$, Co$_3$O$_4$ and CoMn$_2$O$_4$ with varying Co valences.*

In addition to the spin-orbit coupling, multiplet splitting is key to analyzing XPS data. Multiplet splitting arises from the coupling of angular momentum of two electrons in different orbits[67]. If an atom has an unpaired core electron and an unpaired valence electron, the two electrons momentum can couple and result in multiple peaks. This coupling follows the *L-S* or Russell-Saunders coupling scheme. The peaks are separated based on the total angular momentum, *j*, of the coupled electrons. Spin-orbit coupling depends only on the single photoelectron emitted and thus is consistent and easily predictable effect. In the case of multiplet splitting however, the splitting depends on the both the core electron and valence electrons in the same atom. Additionally, given the presence of multiple unpaired valence electrons, the core electron can pair with any of them and each could yield a different energy shift. Most of the coupling, however, will be of relatively low intensity and those that are not are often well documented. Multiplet splitting is commonly observed in transition metal ions that have partially filled *d* orbitals. For example, the valence of Mn can be determined from the multiplet splitting of the Mn 3s core level[68], while different valences and atomic coordinations of Cr exhibit differing multiplet features[69].

Though there are several physical mechanisms that drive their presence, satellite peaks represent an important source of information to interpret XPS data. These peaks appear on a spectrum near a primary core-level peak that are created by electrons from the same conditions as those of the major peak. Shake-up satellite peaks occur when an atom is photo-ionized, which leads to a resulting in core level hole and subsequent electronic rearrangement[70]. In a "shake up" shift, an electron in the valence level becomes excited to a higher state leaving the atom in an excited state. In a "shake down" shift, a core level electron in a higher level might fall down moving the hole up. Both of these alternate states have an energy difference from the ground state the ion would otherwise be in. This energy difference comes out of the kinetic energy of the photoelectron and results in a satellite peak that appears to have a different binding energy than the major peak. For example, Co$^{2+}$ ions exhibit a large shake-up satellite peak at higher binding energy[71], as can be seen in Figure 3.

The third common source of satellite peaks come from charge transfer between atoms. When an atom is photo-ionized, it may result in a change of the electrical structure of itself and an adjacent atom. This change in ionization and electron configuration causes a change in the potential photoelectrons feel. This is apparent from the measured kinetic energy of the photoelectrons on an XPS spectrum. Additional satellite created by charge transfer can appear as shoulders on primary peaks or as completely distinct peaks a few eV away from the primary. An important thing to note is that both the $j = 1/2$ and $j = 3/2$ each have a corresponding satellite and that the ratio of intensity of 2 : 1 is still maintained for the satellites. Charge transfer satellites are observed in Ti 2p and O 1s spectra for materials such as $SrTiO_3$[72].

Shake up and charge transfer satellite peeks should not be confused with the different valence peaks. In ionic systems such as oxides, changes of the valence state of ions is expected. When ions lose or gain valence electrons, the potential of the atom changes as well. This change is typically on the order of a few eV and different valence states often do not result in independent peaks but appear as the convolution of multiple valence peaks, including their component satellite features. This can be a very powerful tool to identify the bonding environment of atoms in a material, but also makes data analysis significantly more complicated than is often incorrectly presented in the literature. As an example, we show three distinct spectra for different Co valences acquired in our lab in Figure 3. Naïve fits to the data will generally not be sufficient to accurately determine the ratio of $Co^{2+}$ to $Co^{3+}$ in a mixed valence material. Instead, readers are encouraged to refer to literature guides for rigorous fitting procedures[66]. The Co $L_3M_{23}M_{45}$ Auger electron also generates peak intensity at ~777 eV, accounting for the shoulder at low binding energy below the Co $2p_{3/2}$ peak, which can complicate quantification and peak fitting.

### C. Integration of XPS with Thin Film Growth (0.5 pages)

We will focus on various examples of the experimental benefits of *in situ* (or more appropriately *in vacuo*) XPS studies in later sections. However, it is first worth considering the practical aspects of constructing an integrated system that combines film growth and surface science studies. Several vendors provide turn-key platforms for research labs that offer MBE or PLD synthesis chambers connected by vacuum transfer to surface analysis chambers that may include some combination of LEED, XPS, ARPES, and STM. These integrated platforms provide remarkable new capabilities for a research lab but are also beyond the start-up budget for new researchers at many institutions. However, through careful design it is feasible to construct an integrated system from modifications to existing systems or by connecting systems virtually through a vacuum suitcase apparatus. In the case of our Auburn laboratory, we combined a newly-purchased MBE system (Mantis Deposition) with a refurbished XPS (PHI 5400 refurbished and resold by RBD Instruments) by repurposing existing vacuum hardware and considering possible future expansion in our design. The integrated chambers are shown in Figure 4. We offer some advice here for others who may pursue similar projects in the future.

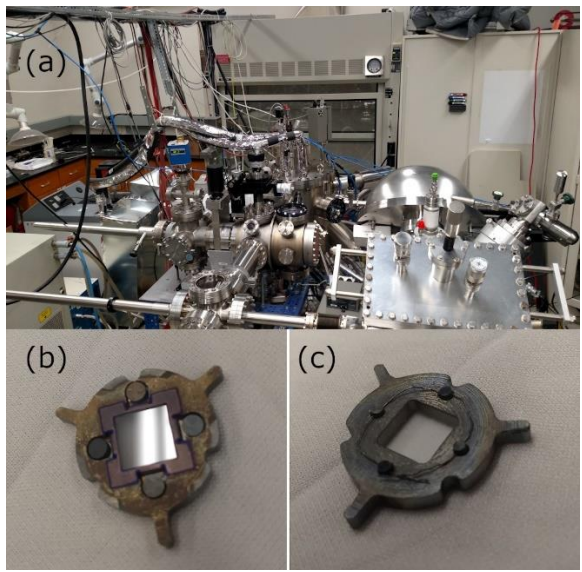

*Figure 4: (a) Integrated MBE (top) and XPS (bottom) system at Auburn University; (b) Front and (c) Back of custom sample holder designed for ease of transfer into XPS.*

In the case where groups are attempting to construct an integrated system on a budget by making use of existing or refurbished equipment, significant care must be taken to make sample transfer practical. For example, the default design for many MBE sample holders makes use of a gravity-held approach, so that the sample cannot be rotated upside-down in vacuum. During design of the MBE system, we worked with the manufacturer to create a custom sample holder that would be practical for high temperature growth in an oxygen environment via backside radiant heater and be transferable to an initially-unknown XPS chamber where it would need to be flipped over using a manipulator. An additional concern was the likelihood that the sample would have to pass through a small 2.75" conflat port to enter the XPS chamber. This design concern necessitated a reduction in the size of the sample holder to a diameter of ~1" and corresponding adjustments to the MBE sample stage to accommodate the smaller holder. The holder is shown in Figure 4(b-c). To our knowledge, this design is completely unique, but offers a great deal of flexibility for groups who may wish to integrate existing chambers in the future. Custom manipulators were designed and machined for each step in the transfer process. We also note that care must be taken to ensure that the vacuum chamber mounts are designed in such a way that the chambers can be coarsely aligned to connect transfer lines. In our case, we picked a fixed height for the MBE chamber and ensured that all other chambers would be mounted on threaded rods or adjustable feet to level the chambers within a few mm of tolerance. The remaining misalignment can be compensated using port aligners and bellows.

Design of the transfer chamber is important for both practical day-to-day use and for quality of data generated in the XPS. To preserve good base pressure in the XPS, it is best for the transfer chamber to have a base pressure of <$10^{-8}$ Torr. This means that a loadlock that is isolated from the transfer chamber is strongly preferred. Ideally, transfer between the MBE and XPS can occur within a few minutes after removing the sample from the growth chamber so that multiple samples can be synthesized and measured in a single workday. Additionally, by reducing the background pressure and time in the transfer chamber, water adsorption on the sample can be reduced for better XPS data. Designing a storage stage in the transfer chamber or load-lock that can hold additional samples for future XPS measurements is desirable so that XPS measurements do not become a bottleneck to overall productivity in the MBE.

# II. Combining Growth and XPS to Measure Film Quality
## A. Measuring Contaminants

Epitaxial thin film research has a very low tolerance for elemental contamination. Despite engineering controls, contaminants from unknown sources sometimes make their way into samples and can either hinder data acquisition in experiments or change the physical properties of the material entirely. Surface-sensitive XPS is an excellent way to detect any unwanted element contamination, with an elemental sensitivity of <1 atomic percent. The most efficient way to measure contamination is by acquiring wide-range surveys of samples from maximum binding energy to zero. In general, it is good practice to perform surveys on every sample before performing a longer-duration, high-resolution measurement over core level and valence band regions. XPS surveys allow for a quick qualitative depiction of a sample's surface stoichiometry in addition to indicating any unexpected elemental contaminants.

As an example, Figure 5 is a survey of a $CoMn_2O_4$ spinel thin film sample grown in our MBE system. There are many peak features in the survey representing cobalt, manganese, and oxygen signal, but additional peaks indicate a contaminating element. Based on the position and shape of these extra features, we determined the presence of sodium on the sample surface. By inspecting surveys of samples grown at different conditions, we pinpointed the source of the sodium to the radio-frequency oxygen plasma source. After discussions with the manufacturer, the Na-contaminated quartz discharge tube was replaced with an alumina discharge tube that has eliminated contamination. We are aware of at least one other plasma source that has also been a source of Na contamination and suggest that groups should be careful to test new plasma sources upon purchase. Other contaminants can be detected by XPS, such as F due to surface treatments performed on $SrTiO_3$ with a buffered oxide etchant[73]. In this sense, the availability of *in situ* XPS can accelerate the calibration process for ideal growth conditions and pristine materials, making it valuable as a day-to-day diagnostic tool even when the data will not be used for publishable research.

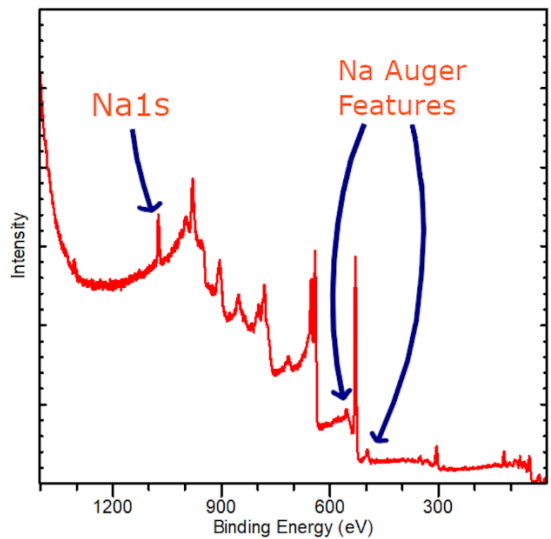

*Figure 5: XPS survey of $CoMn_2O_4$ thin film showing Na contamination from oxygen plasma source.*

## B. Film stoichiometry and surface termination

Because of the nature of photoelectron scattering within the crystal prior to escape, XPS measurements are inherently surface sensitive. For a given atomic species with volume density ρ, the measured intensity as a function of emission angle θ relative to the surface normal is given by:

$$I(\theta) = I_0 \rho(z) \sigma_{n,l} T(E_k, \theta) A \int_0^\infty e^{-\frac{z}{\lambda \cos(\theta)}} dz,$$

Where $I_0$ is the intensity for $\theta = 0°$, $\sigma_{n,l}$ is the absorption cross section for a given energy level with principal quantum number $n$ and orbital angular momentum $l$, $A$ is the measured area on the sample, $z$ is the depth within the material, $T(E_k, \theta)$ is the analyzer transmission function, and $\lambda$ is the inelastic mean free path (IMFP). This equation is generally simplified by assuming that the transmission function is constant over the energy values of interest and by using ratios between core level intensities to eliminate the role of the absorption cross sections, $I_0$, and $A$. For more detailed descriptions of the physics of depth sensitivity in XPS, the reader is referred to the work of Powell and Tanuma[74].

In practice, the depth sensitivity of XPS can be both a blessing and a curse. The angle-resolved XPS technique can be used to measure the surface termination of a sample, cation intermixing across an interface, and determine a depth profile of specific chemical features, providing rapid new depth-resolved insights into synthesized materials without destroying the sample through sputter etching or focused ion beam liftout for electron microscopy measurements. Conversely, however, the surface sensitivity of XPS means that a change as small as flipping the surface termination from $TiO_2$ to SrO on a $SrTiO_3$ single crystal can reduce the measured Ti:Sr peak area ratios by ~10%. This makes stoichiometry quantification for epitaxial thin films by XPS very difficult, with conventional sensitivity factors providing very little value. It is generally best to benchmark the area ratios with a bulk sensitive technique such as Rutherford back scattering or a single crystal reference and compare subsequent films to the measured area ratios from the calibrated reference sample while taking care to only compare samples with known surface terminations.

However, despite the challenges of absolute stoichiometry quantification via XPS, we have found a convenient trick that employs *in situ* capabilities to accelerate the calibration process for oxide MBE growth. Given the layered nature of the perovskite crystal structure and the propensity for excess cations to reside on the film surface[75,76], using angle-resolved XPS to determine the surface termination is generally an effective means to determine whether a perovskite oxide film has excess *A*-site or *B*-site cations even without an RBS or single crystal reference standard.

Beyond the convenience of angle-resolved XPS in the synthesis process, it has helped to explain surface-dependent phenomena in a variety of systems. We have employed angle-resolved XPS to determine for the first time that an SrO termination is more stable for stoichiometric $SrTiO_3$ films grown by hybrid MBE[77]. Furthermore, in the case of a $LaAlO_3/SrTiO_3$ heterostructure grown on SrO-terminated $SrTiO_3$, an unexpected $AlO_2$ termination was observed by angle-resolved XPS, which helped to explain the absence of conductivity at the interface[78]. Angle-resolved measurements of $LaFeO_3$ film surfaces have also helped to explain the role of surface termination on chemical reactivity for catalytic water splitting[79]. Despite the simplicity of angle-resolved XPS and the relatively mundane knowledge gleaned from knowing the surface termination of a thin film, there is a great deal of new physical understanding that can arise from such day-to-day measurements through *in situ* XPS.

### C. Oxide Surface Chemistry

While understanding of the surface termination regarding perovskite AO or $BO_2$ layers is valuable for film synthesis, it only begins to address the subtleties of surface chemistry that occur in complex oxides. Measurement of the O 1s peak can also provide valuable information to understand how the surface is passivated, a process that occurs in different ways depending on the surface polarity. It is known that water will adsorb differently depending on the AO or $BO_2$ termination of a (001)-oriented surface[77,79]. However, even in UHV after *in situ* growth, the adsorption of water on the surface can impact the electronic properties of the material. Studies that integrate XPS with atomically-resolved scanning tunneling microscopy are incredibly valuable in this regard to demystify some of the surface phenomena that drive emergent behavior.

As discussed above, the KTao$_3$ surface 2DEG is of significant interest for the Rashba splitting that is observed[33]. Through STM and XPS studies, Setvin et al. demonstrated that a 2x1 surface reconstruction of the surface with uniform K(OH)$_2$ coverage after water exposure[38]. An OH peak on the O 1s core level at higher binding energy was reported for this case to confirm their results. Similar experiments have also been performed on layered ruthenates (Ca$_3$Ru$_2$O$_7$[80] and Sr$_2$RuO$_4$[81]) and show that water can adsorb on the surfaces of oxides in a variety of different configurations with different binding energies relative to the metal oxide O 1s peak that is typically at ~530.0 eV. It is generally observed that films with AO termination will more readily adsorb water[38,77,79,81], though the exact surface structure is very difficult to decipher. The effect of the adsorbed water—even in situ after growth—should not be discounted when analyzing the observed properties. For example, it is common in the literature to associate a higher binding energy peak 2.3 eV above the metal oxide binding energy with oxygen vacancies, but the origin of this peak is the subject of some controversy and it can easily be due to other surface chemistry effects that will confuse the analysis.[79,82]

## III.   Probing Interfacial Phenomena in Oxide Heterostructures with XPS

### A.   Interfacial Chemistry

The surface sensitivity of XPS through angle-resolved measurements can be extremely useful for the purposes of determining depth-dependent phenomena in oxide heterostructures. We have already showed the effects of surface variations and how they can be used to understand oxide film growth. However, the same techniques can be taken a step further to understand such behavior as interfacial intermixing in heterostructures and growth-induced defects such as oxygen vacancies. By varying the photoelectron emission angle to the detector and modeling the angular dependence of the relative peak areas based on the IMFP for the sample, one can construct a quantitative model of the cation or defect profile as a function of depth within a material.

For a buried interface, angle-resolved measurements can be used to model the depth of defects or dopants. The approach has been particularly valuable in studies of polar/non-polar interfaces, where differences between the idealized physical model of an interface and the actual synthesized heterostructure have led to confusion as to the physical origin of emergent behavior. In the case of LaAlO$_3$-SrTiO$_3$ interfaces, angle-resolved measurements have shown that cations from the underlying substrate will readily out-diffuse into the LaAlO$_3$ film, with the degree of diffusion dependent on the cation stoichiometry of the LaAlO$_3$ film[19,83]. These measurements set the stage for further studies that confirmed that the 2-dimensional electron gas at these interfaces is dependent on the film stoichiometry.[20,21] Other angle-resolved studies have shown that oxygen vacancies in SrTiO$_3$-based heterostructures are localized at the interface[84,85].

In the case of a SrTiO$_3$-LaCrO$_3$ superlattice, *in situ* angle-resolved XPS measurements were used to determine the degree of Cr out-diffusion into the topmost SrTiO$_3$ layer[86]. The resulting data was compared to a model with differing degrees of Cr intermixing to estimate the Ti-Cr concentration profile across the B site of the superlattice. The model agreed very well with extracted concentration profiles from STEM-EELS measurements, as shown in Figure 6[86].

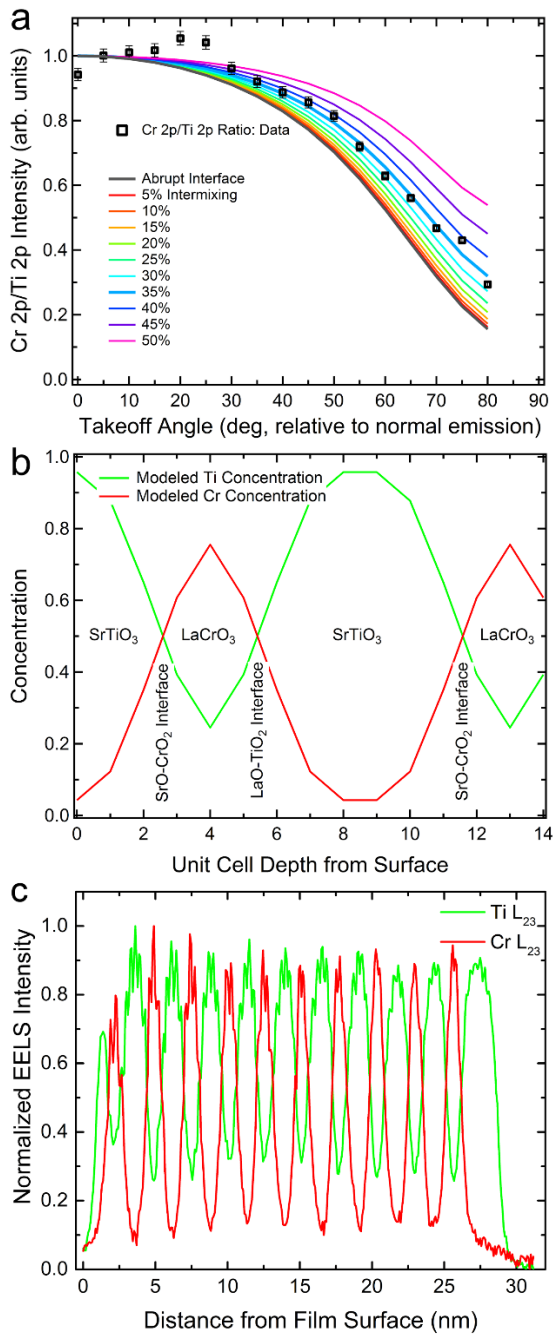

*Figure 6: a) Angle-resolved XPS Cr 2p:Ti 2p peak ratio with models assuming various degrees of intermixing (0° deg is normal to the film surface); ; b) Concentration profile for 35%-intermixing model; c) STEM-EELS integrated signal profile throughout superlattice determined using MLLS fitting of the Cr $L_{23}$ edge and the background-subtracted peak area of Ti $L_{23}$ edge. The signal has been normalized to the substrate. Reprinted with permission from Comes et al, Chem. Mater. **29**(3), 1147. Copyright 2017 American Chemical Society.[86]*

## B. Band Alignment and Potential Gradients

Semiconductor heterostructures are key to modern-day electronics and numerous oxide heterojunctions have been pursued for their potential applications. Performance of these heterostructures can be engineered by manipulating valence and conduction band offset at the interface. It is crucial to understand this band offset to accurately predict and manipulate the behavior of thin film heterostructures.

There are different schemes to determine the band offset experimentally such as ultra-violet spectroscopy, internal photoemission spectroscopy, and XPS. Band offset determination via XPS has been used extensively. Unlike other methods, XPS is sensitive to details at the interface such as changes in valence and chemical intermixing and thus is often preferred over other methods. The XPS approach to valence band alignment determination was first introduced by Kraut *et al*[87]. It is based on the premise that the energy difference between a core level and valence band maximum (VBM) is an intrinsic property and remains constant independent of any formation of a heterostructure. Any change in VBM will change the core level equally, making the energy of the core level peak a good proxy for the VBM. This becomes important when measuring a thin film heterostructure, as the valence band region will now be the convolution of two or more distinct materials, while appropriately chosen core levels can still be measured easily and fit repeatably. The precision of this method lies in the measurement of core level and valence band maximum for a single crystal reference or carefully grown thick film. In this case, a thick film is one that is sufficiently thick such that the underlying substrate does not contribute any signal to the data. Using their data, Kraut *et al.*[87] determined the VBM by fitting valence band spectra with the theoretical valence band density of states. However, fits involving a linear extrapolation of the leading edge of the valence band-spectra to the zero-level backgrounds have been found to be as accurate and require much less effort. Various refinements to this approach over the years for oxide heterostructures have been proposed[88,89]. A diagram illustrating Kraut's method is shown in Figure 7(e). The valence band and conduction band offset are calculated as

$$\Delta E_v = \left(E_{CL}^A - E_V^A\right) - \left(E_{CL}^B - E_V^B\right) - \left(E_{CL}^A - E_{CL}^B\right)$$

$$\Delta E_C = E_g^A - E_g^B + \Delta E_V$$

The core levels peaks and valence band reference spectra should be acquired with high energy resolution, as the uncertainty of the measurement of the core level peak position will propagate into the band offset calculation. Choice of peaks that have a narrow intrinsic width and are easier to fit repeatably is thus very important.

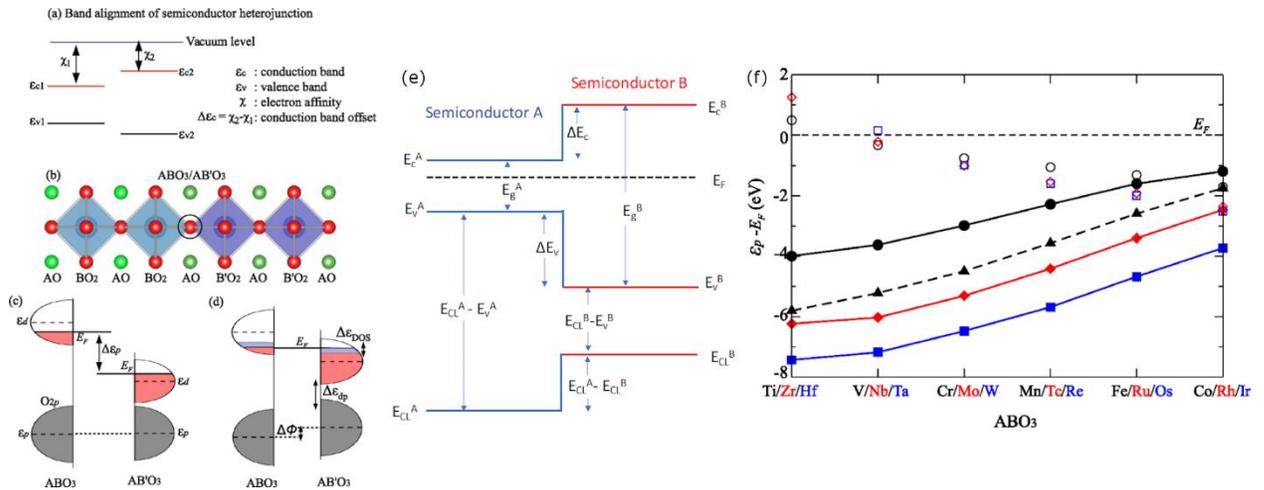

*Figure 7: (a) Conventional model for semiconductor band alignment based on aligning the vacuum level for constituent materials; (b) Schematic model of $ABO_3/AB'O_3$ interface emphasizing continuity of O 2p electronic states; (c) Preliminary band alignment based on alignment of O 2p states prior to charge transfer; (d) Reconstructed band alignment after charge transfer equilibration of the Fermi energy level; (e) Schematic rendering of method to measure band alignment in XPS via core level binding energies; (f) Summary of bulk $\epsilon_p$ (filled symbols) and $\epsilon_d$ (empty symbols) with respect to the Fermi level ($E_F=0$) for different $SrBO_3$ (solid line) materials, with 3d (black), 4d (red), and 5d (blue) elements. The simple criterion for the direction of*

*the charge transfer at the ABO$_3$/AB'O$_3$ interface is that the component with lower (more negative) $\epsilon_p$ will donate electrons to the other one. Also plotted is LaBO$_3$ (dashed line) for B=3d, for estimates for ABO$_3$/A'BO$_3$ interfaces. Adapted under Creative Commons Attribution 3.0 License from Zhong and Hansmann, Physical Review X, **7**, 011023 (2017).[90]*

The LaAlO$_3$-SrTiO$_3$ interface has been widely studied for its unusual transport properties. The properties can be attributed to an effective 2-dimensional electron gas (2DEG) at the interface. The origin of this 2DEG is not known exactly but there exists some hypotheses for its origin[91]. Owing to its high precision in measuring core-levels and VBMs, and subsequently valence band and conduction band offsets, XPS has been a powerful tool in studying this heterostructure interface. Additionally, XPS data can be modeled to estimate the potential gradient across the film.

Multiple groups have used XPS *to* study the LAO-STO interface grown by both MBE[22] and PLD[19,92,93] with different film thicknesses and terminations for possible explanations for the origin of the interface's 2DEG. In one such work, Segal et al. confirmed the metal-insulator interface transition with 4 unit cell (UC)-thick LAO on a TiO$_2$ terminated surface.[22] Many studies suggest that the 2DEG at the interface stems from an electronic reconstruction due to polar discontinuity (LAO has alternating positive and negative atomic layers), known as the "polar catastrophe."[94]. From the experimental band offset and band gap, *Segal et al.* calculated that a potential gradient of 1.15+/-0.06 at a 4 UC thickness is required for the polar catastrophe to occur, whereas the measured potential gradient at 4 UCs was much less than the required value. This means that other phenomena such as oxygen defects, lattice distortion and cation mixing at the interface must be investigated to explain this unique behavior at the interface. In this way, XPS has helped to further elucidate the origin of 2DEG at the LAO-STO interface, which was later shown to be strongly dependent on LaAlO$_3$ cation off-stoichiometry[20,21].

Other studies of interfaces where both sides of the heterojunction are comprised of band insulators with the valence band maximum comprised of O 2p-derived electronic states have generally shown small valence band offsets of ~0.5 eV or less[95]. This includes the SrTiO$_3$-(La,Sr)(Al,Ta)O$_3$ interface[96], the SrZrO$_3$-SrTiO$_3$ interface[97], and the BaSnO$_3$ interface with both SrTiO$_3$ and LaAlO$_3$[98]. The small valence band offset can generally be attributed to the condition that the O 2p energy level be continuous across the heterojunction,[90] which will be discussed in more detail shortly.

Unlike the isoelectronic interfaces where O 2p states form the top of the valence band on both sides of the heterojunction, interfaces between Mott-Hubbard and band insulators offer additional degrees of freedom. For example, the LaFeO$_3$/*n*-SrTiO$_3$ system also creates a polar non-polar interface and is studied for its photocatalytic applications[99–102]. M. Nakamura *et al.*[103] studied the interface-induced polarization of LaFeO$_3$/Nb-doped SrTiO$_3$ heterostructures and observed novel shift current that suggested the polarization direction switches for different substrate terminations (SrO and TiO$_2$). This would mean that potential gradients took on opposite signs for different terminations and suggested that changes in polarization is responsible for observed photoconductivity. Subsequent studies of LaFeO$_3$/Nb-doped SrTiO$_3$ junctions with varying LaFeO$_3$ film thicknesses (3,6 and 9 UCs) and different substrate terminations (SrO and TiO$_2$) grown using MBE and characterized by *in situ* XPS helped to address these questions[101]. These studies found an increase in separation between core levels (La 4d and Sr 3d) with increases in film thickness, clearly indicating that band offset and built in potential changes with thickness. In addition, by modeling the broadening of the La 4d peaks with different film thicknesses and terminations, it was possible to estimate the potential gradient across the film. Contrary to the reported interface-induced polarization[99,100], the results indicated that the interface termination had only a small impact on the band alignment and the built-in potential gradient[101]. The models are shown in Figure 8. Chemical instability of the SrO/FeO$_2$ interface has been proposed as a likely explanation for these results[104]. These results were subsequently matched by first-principles band alignment models of the interface[102].

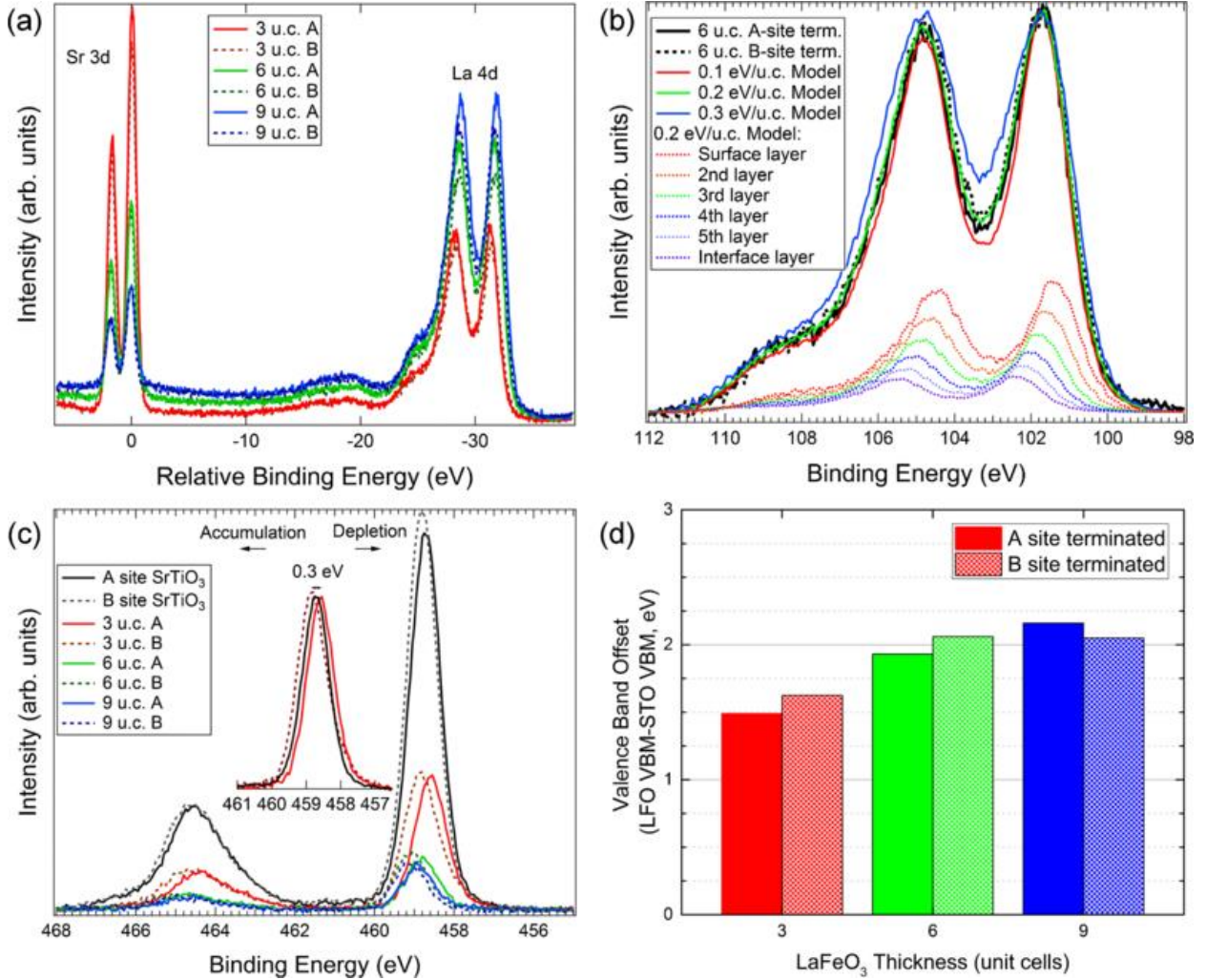

*Figure 8: (a) Sr 3d and La 4d core-level spectra for the family of heterostructures, shifted to align the Sr 3d peaks; (b) model of La 4d peak broadening in the 6 u.c. films; (c) Ti 2p core-level spectra for each film and substrate, with the inset showing the peak shifts; (d) valence band offsets determined from the core-level spectra for each heterojunction. Reprinted figure with permission from Comes and Chambers, Physical Review Letters, **117**, 226802 (2016).[101] Copyright 2016 by the American Physical Society.*

### C. Interfacial Charge Transfer

As in modulation-doped semiconductor heterostructures, interfacial charge transfer is an important tool in complex oxides to induce novel magnetic and electronic functionalities that do not occur in equilibrium. One of the crucial parameters to determine the charge transfer probability is the charge transfer gap. In most of the cases the charge transfer gap is determined by band splitting of B site cations or the band gap between B site cations and fully occupied O 2p bands, while A site cations have a less significant role. Predicting charge transfer is not straight forward in oxide heterostructures since the classic band alignment principle is not strictly followed as in the case of semiconductors. A modified rule is based on the continuity of states of the O 2p band across the interface and allows for a qualitative prediction of band alignments and charge transfer in complex oxide heterostructures[90]. In cubic oxides, the local energy of oxygen 2p states, $\mathcal{E}_p$, is the deterministic factor to tune the direction of charge transfer from one TMO to another. Generally, electron transfer favors the direction from lower $\mathcal{E}_p$ to higher. The classic rule is shown in Figure 7(a), while the modified rule for oxides is shown in Figure 7(c-d). This rule led to the development of alternative computational methods to predict charge transfer that have been shown to be effective in recent years.

Density functional theory (DFT) is one of the most widely used theoretical modeling techniques to predict the charge transfer mechanism in metal oxides interface. Different metal oxides interfaces follow different charge transfer mechanisms. Particularly in Mott-Hubbard insulators, the O 2p band alignment and the competition between crystal field and correlation energy of d electrons, are crucial to determine the electronic rearrangement[90]. Generally, the apical oxygen atom at the interface (see Figure 7(b)) is shared by the materials at the heterostructure interface. This results in the alignment of O bands at the interface. However, considering the O band alignment alone disregards the creation of the internal electric field that balances the electrochemical potential between two B-sites and prevents further charge transfer. Hence the interfacial charge between two materials transfer cannot completely be understood relying only on O 2p band alignment. An extra factor that comes into play is the rearrangement of the d- bands on B-site cations. Specifically, the density of states of B-site cations of each material reflects a clear picture of possible electronic rearrangement.

As an example, the Ti atom would be expected to donate electrons to the Fe atom at the $LaTiO_3/LaFeO_3$ interface, implying single a formal charge of 4+ for the interfacial Ti ion and the shifting of Fe from a 3+ state to a 2+ state near the interface[90]. A theoretical prediction of charge transfer in metal oxides interface is supported by XPS analysis both qualitatively and quantitatively to resolve the valence band structure. XPS is well known for its high sensitivity to the variations in valence states of transition metal ions. J.E. Kleibuker *et al*[105] showed an extra peak at ~2eV lower binding energy is observed in LFO/STO heterointerface in reference to the bulk LFO for Fe 2p XPS spectra, as shown in Figure 9. This extra peak in Figure 9(c) at slightly lower binding energy highlighted by an open circle represents the presence of $Fe^{2+}$ state along with $Fe^{3+}$ state at higher binding energy denoted by a solid circle. Interestingly, the extra peak gets more pronounced with the decrease in thickness of the $LaFeO_3$ layer which represents the charge transfer is taking place near or at the interface. To understand the rearrangement mechanism for transfer of charge, a valence band spectrum in XPS is analyzed experimentally. The evolution of an extra peak around 1 eV in valence band spectra represents the completely-filled $t_{2g}$ band of $Fe^{2+}$ in a low-spin configuration. Separate measurements of the Ti 2p spectrum confirmed 4+ formal charge of Ti, indicative of charge transfer from Ti to Fe of 1 $e^-$/unit cell at the interface.

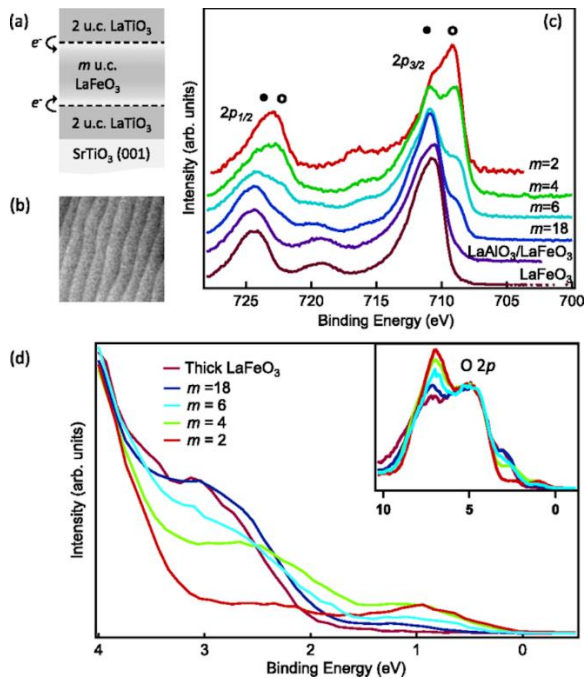

Figure 9: (a) Sketch of the $LaTiO_3/LaFeO_3$ sample geometry. (b) A typical 1×1 μm AFM height image of a $LaTiO_3/LaFeO_3$ heterostructure. (c) Fe 2p XPS spectra of $LaTiO_3/LaFeO_3$ heterostructures for various thicknesses of $LaFeO_3$, as well as of a

30 u.c. LaFeO$_3$ film and a (2/2) LaAlO$_3$/LaFeO$_3$ heterostructure. The solid and open circles mark the Fe$^{3+}$ and Fe$^{2+}$ peaks, respectively. (d) Valence band XPS spectra of LaTiO$_3$/LaFeO$_3$ heterostructures for various thicknesses of LaFeO$_3$. All spectra were taken near normal emission ($\theta=3°$). Reprinted figure with permission from J.E. Kleibeuker et al, Physical Review Letters, **113**, 237402 (2014).[105] Copyright 2014 by the American Physical Society.

For other instances, preparation of complex superlattices that are different from regular heterojunction helps to enlighten the process of charge transfer. One such example is comparison of LaTiO$_3$/LaNiO$_3$/LaAlO$_3$ with a LaNiO$_3$/LaAlO$_3$ interface where the Ni atom has doubly degenerate $e_g$ manifold with one electron. In the bulk structure, B- site cations of each material have 3+ oxidation state. As Ti has 4+ as its most stable oxidation state, one electron can move from Ti $t_{2g}$ band to Ni $e_g$ band resulting Ni transition to a 2+ state from a 3+ state[5]. The insertion of the LaAlO$_3$ layer induces symmetry breaking and polar structural distortion that contributes to the phenomenon of orbital polarization. X-ray absorption spectroscopy (XAS) confirmed this charge transfer and also showed a preferential orbital occupation of the Ni $3d_{z^2}$ $e_g$ orbital, producing the orbital occupation that makes LaNiO$_3$ resemble the superconducting cuprates. Similar measurements in a LaNiO$_3$/LaMnO$_3$ heterostructure indicated electron transfer from Mn to Ni atom at the interface[4]. Finally, a combined XPS and XAS study of LaCoO$_3$/LaTiO$_3$ heterostructures also indicated charge transfer to produce Co$^{2+}$ and Ti$^{4+}$ valences at the interface[106]. Each of the examples we have shown here agree with the model put forward by Zhong and Hansmann[90], indicating the value of their approach in predicting interfacial charge transfer and band alignment. Future studies may benefit from moving beyond 3d transition metal systems into the 4d and 5d blocks of the periodic table to generate novel interfacial phenomena.

### D. Experimental Design Considerations

We have shown above several examples of the use of *in situ* XPS studies to probe interfacial phenomena in oxide heterostructures. The availability of an XPS appended to a growth chamber makes these studies very convenient and generates important physical insights as soon as the synthesis is complete. However, the knowledge gleaned from such studies is constrained by both the physical limitations of XPS and by the researcher's chosen design for the synthesized heterostructure. We have already discussed the information depth within the sample for an Al K$\alpha$ source, which is limited to ~5 nm for photoemission normal to the film surface. Thus, band alignment and charge transfer studies are impractical for interfaces that are more than 5 nm below the surface.

One must also consider which interfaces in a multilayer will exhibit emergent properties, the chemical stability of the surface in vacuum and in atmosphere, and potential interference due to overlapping core level peak energies. In the latter case, LaNiO$_3$ is a good example of a material that cannot be easily examined via XPS due to the overlap of the La $3d_{3/2}$ and the Ni $2p_{3/2}$. The LaNiO$_3$/LaTiO$_3$ charge transfer described above would be very difficult to perform in XPS, though use of the Ni 3p peak would provide one way to probe changes in Ni valence[107]. Furthermore, the depth sensitivity of XPS means that if two interfaces are present in a multilayer, as in the case of the LaTiO$_3$/LaFeO$_3$/LaTiO$_3$ trilayer in Figure 9(a), then the top interface will generate a majority of the signal and any phenomena at the bottom interface may be difficult or impossible to measure. Finally, while *in situ* measurements protect against changes due to atmospheric exposure, some surfaces may be modified even during the sample cooldown process[32] or will have intrinsic oxygen vacancies[55]. Deposition of a protective capping layer during the growth process that will not induce any chemical changes in the underlying materials is thus valuable even for *in situ* studies if a surface is unstable.

## IV. Future Directions for *In Situ* and Related Studies
### A. Hard X-ray Photoelectron Spectroscopy

Advances in spectroscopy over the past decade have largely been driven by the development of new techniques from next-generation synchrotron light sources. These have included the emergence of hard X-ray photoelectron spectroscopy (HAXPES). A variety of new HAXPES beamlines[108–111] have been constructed as these light sources are brought online, enabling new XPS studies that are not limited by the surface sensitivity of lab sources. Performing XPS studies with X-ray photon energies >2 keV provides more bulk sensitivity and enables access to transition metal 1s core levels that cannot be examined in a conventional Al Kα laboratory source. As discussed previously, the inelastic mean free path (IMFP) for photoelectrons, λ, scales with the square root of electron kinetic energy for high kinetic energies. This means that electrons with kinetic energies ~10 keV will have IMFP approaching 10 nm. An IMFP of 10 nm produces depth sensitivities such that only half of the escaping photoelectrons will be from the top 6.5 nm of the sample and only 10% of the signal comes from the surface (top 1.0 nm). Conversely, for an IMFP of 1.5 nm that would be typical for a lab source, half of the signal is generated from the top 1.0 nm and only 3% of the signal comes from 5.0 nm below the surface. Thus, HAXPES measurements enable studies of deeply buried interfaces more than 5 nm below the film surface that would be impossible with a lab source. Numerous synchrotron studies have made great use of HAXPES to probe oxide interfaces, and we refer the reader to several good reviews of these studies.[112–115]

As an example of the benefits of HAXPES and the opportunities for further growth, we examine several studies of $BaSnO_3$ thin films and heterostructures. $BaSnO_3$ is an exciting wide bandgap oxide that has been shown to exhibit extremely high electron mobilities at room temperature (150-200 $cm^2$/V-sec) when grown by molecular beam epitaxy.[116–118] Unlike other perovskite oxides, the conduction band of $BaSnO_3$ is derived from Sn 5s orbitals,[119,120] which enable a much greater degree of electronic dispersion and carrier mobility than i.e. 3d orbitals in $SrTiO_3$. While this dramatic enhancement of mobility is of great benefit for potential future device applications, such as field effect transistors[121–123], it also poses challenges to effective doping, both through the use of *n*-type donors[124] and across interfaces through modulation doping and charge transfer[98]. HAXPES has been particularly useful to elucidate the nature of the electronic transport in $BaSnO_3$ and in examining interfacial carrier profiles in heterostructures.

Initial studies of La-doped $BaSnO_3$ via HAXPES focused on understanding the nature of the conduction band profile in the material, by probing electronic states near the Fermi level[119,120]. Through *ex situ* synchrotron HAXPES studies with 4 keV photon energy, the first evidence of mobile Sn 5s electrons near the Fermi level was reported[119]. Laboratory-source Al Kα studies of the same samples could not confirm the presence of carriers at the Fermi edge. This observation was attributed to the greater photoelectron cross section for Sn 5s with 4 keV photons in comparison to the 1487 eV lab-source. However, given continued studies of the stability of dopants in $BaSnO_3$[124], it also seems possible that atmospheric exposure led to depletion of the surface carriers, which would explain the absence of observed carriers in the lab-source experiments. Carriers in identically doped $BaSnO_3$ were later observed in a lab-source experiment[125], lending further credence to this possibility. It is also possible that the HAXPES experiments[119] were successful due to the greater probe depth for hard X-rays, which overcomes the surface depletion. Further HAXPES studies of $BaSnO_3$ doped with differing La concentrations helped to explain the nature of the conduction band filling and band gap renormalization due to the Burstein-Moss effect[120]. These studies have paved the way for the development of stannate thin films and heterostructures, as research moves from understanding the nature of electronic conduction to confinement of carriers in a high-mobility 2DEG for device applications.

Modulation doping of $BaSnO_3$ through engineering of band alignment and charge transfer is a key requirement for a 2DEG structure. Impurity scattering from La donor ions is a key limitation to the further enhancements of mobility in electron-doped $BaSnO_3$[117]. To overcome this, engineering of a heterostructure with conduction band electrons at greater energies than the $BaSnO_3$ conduction band is necessary. In one such heterostructure, a 14 nm La-doped $SrSnO_3$ top layer was engineered such that free carriers in that layer could flow "downhill" to an undoped $BaSnO_3$ below the $SrSnO_3$ layer to form a 2DEG[126]. Given the thickness of the $SrSnO_3$ top layer, HAXPES studies were needed to probe the buried interface. Using a photon energy of 5.93 keV, the group was able to extract the interfacial band alignment and estimate the degree of charge transfer into the underlying $BaSnO_3$. These results are shown in Figure

10. Interestingly, the extracted band profile in Figure 10(d) could be explained by surface carrier depletion from atmospheric exposure, as discussed above regarding La-doped BaSnO$_3$[119]. These results show that the band alignment techniques that have been used for *in situ* studies for many years can be applied to more challenging structures using hard X-ray sources that probe more deeply into materials. Continued development of HAXPES will dramatically improve the ability to measure interfacial phenomena in heterostructures that are not limited to a few nm in thickness due to the lab-based X-ray energies.

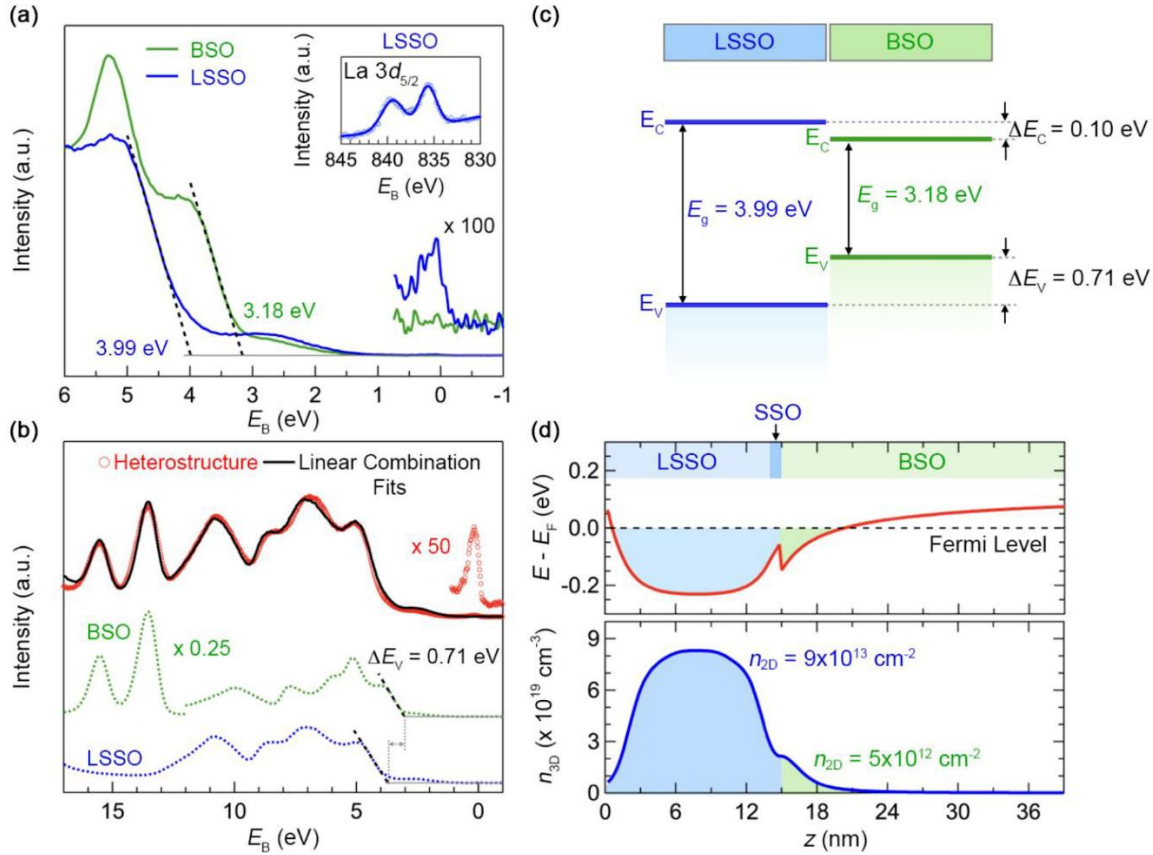

Figure 10: Band alignment at La-doped SrSnO$_3$ (LSSO)/SrSnO$_3$ (SSO)/BaSnO$_3$ (BSO) interface. (a) VB spectra of the reference BSO (green) (56 nm BSO/SrTiO$_3$ (001)) and LSSO (blue) (41 nm LSSO/8 nm SSO/GdScO$_3$ (110)) films. Electronic states near the Fermi states are magnified. Inset shows the La 3d$_{5/2}$ core-level X-ray photoelectron spectra, (b) VB spectra of the SSO/BSO heterostructure (red) along with the fit (black) using linear combination of the reference VB spectra (dotted green and blue lines) to determine the VB offset. (c) Energy-level flat-band diagram showing the measured band offsets between LSSO and BSO, and (d) conduction band minima (red) referenced to the Fermi level (top panel) and 3D carrier density, n$_{3D}$ (blue) as a function of depth for the SSO/BSO (bottom panel). The shaded regions indicate 2D density in LSSO and BSO layers after the charge transfer. Reprinted with permission from A. Prakash, et al, Nano Letters, **19**(12), 8920-8927 (2019).[126] Copyright 2019 American Chemical Society.

### B. Standing-wave XPS

While HAXPES measurements are one approach to probe more deeply within a material and examine both bulk electronic structure and buried interfaces, the underlying physics of the photoemission process is unchanged. As we showed above, HAXPES measurements can still be affected by surface chemistry of the films. The standing-wave XPS (SW-XPS) approach has emerged over the past decade specifically focusing on studies of superlattice structures that exhibit regularly repeating interfaces. By varying the incoming X-ray angle such that the Bragg diffraction condition is satisfied for the superlattice, the intensity of the electric field is modulated across the interface. A stronger Bragg reflection produces a greater standing wave intensity and greater modulation of the electric field strength across the repeating superlattice formula unit. The physics of SW-XPS has been described in more detail in various references[112,127–129]. Here we aim to discuss the applications of SW-XPS to understanding interfacial phenomena and present opportunities to expand on this unique approach by designing interfacial materials that will be well suited for SW-XPS studies.

As an initial example, we present a study of layer-resolved band alignment in a superlattice achieved through careful design of a superlattice structure. The structure and electric field strength modulation is shown schematically in Figure 11 for a study of a $LaCrO_3$-$SrTiO_3$ superlattice using ~830 eV photon energy near the La $M_5$ resonance.[130] In this case, the formula of 10 unit cells of $SrTiO_3$ followed by 5 unit cells of $LaCrO_3$ over 10 repeating layers in the superlattice was chosen to maximize the Bragg reflection and thus the sensitivity to individual interfaces. By choosing the superlattice periodicity prior to growth and considering the subsequent SW-XPS experiment, the largest standing-wave response ever achieved was reported, with individual core level intensities varying by ~50% as the X-ray angle was varied across the Bragg peak. This enabled extraction of the band alignment across the entire 15-unit-cell structure through careful modeling of the core level peak positions as a function of incoming angle, producing good agreement with DFT models[130,131].

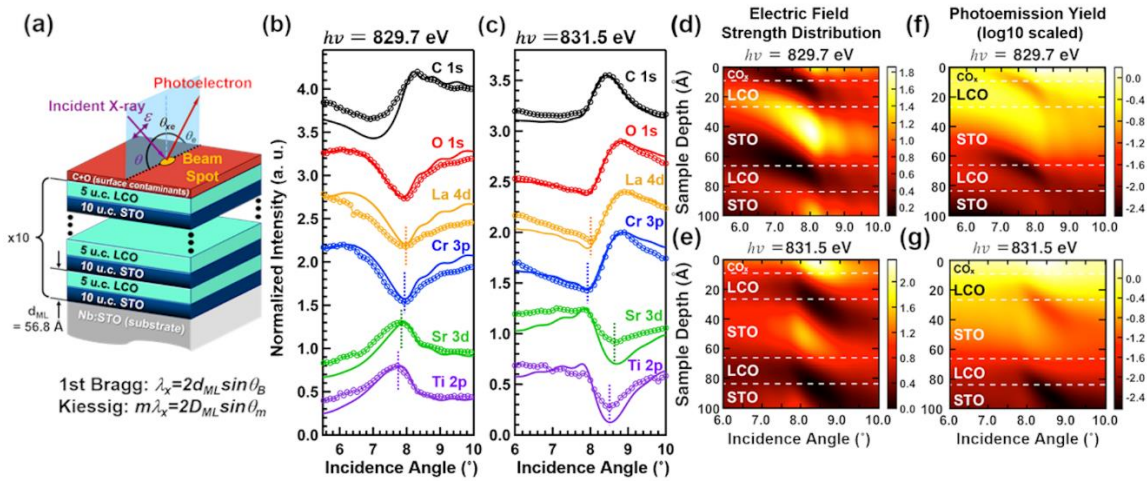

Figure 11: (a) Schematic of the superlattice made up of 10 bilayers of LCO and STO, consisting of 5 unit cells of LCO, 17.6 Å thick, and 10 unit cells of STO, 39.2 Å thick, grown epitaxially on a Nb-doped STO(001) substrate. The two sources of standing-wave structure in the rocking curves are indicated: Bragg reflection from the multilayer with period $d_{ML}$ and Kiessig fringes associated with the full thickness of the multilayer stack $D_{ML}$. Experimental (open circles) and simulated (solid) rocking curves of representative elemental states at photon energies of (b) 829.7 eV, (c) 831.5 eV. The colored dash lines are the guides to the eye to indicate the phase of the rocking curves in (b) and (c) to show sensitivity to the interfacial termination (SrO-$CrO_2$ vs. $TiO_2$-LaO). The electric field strength distribution derived from x-ray optics calculations at two energies near the La $M_5$ resonance, (d) 829.7 eV and (e) 831.5 eV as a function of sample depth and incidence angle. Note the significant shift in position between the two energies. The corresponding photoemission yields, (f) and (g), plotted on log10 scales. Adapted figure with permission from S.-C. Lin et al, Physical Review B, **98**, 165124 (2018).[130] Copyright 2018 by the American Physical Society.

The SW-XPS technique has been used to probe changes in interfacial chemistry in a variety of other superlattice systems. Evidence for $Ni^{2+}$ valence near the interface with SrTiO in 4 u.c. $LaNiO_3$-4 u.c. $SrTiO_3$ superlattices was reported.[132] Studies of $LaNiO_3$-$CaMnO_3$ superlattices showed similar effects, with $Ni^{2+}$ states observed at the interface.[133] Likewise, changes in the interfacial Mn valence in $(La,Sr)MnO_3$-$SrTiO_3$ superlattices were also observed.[134] Collectively, these studies have shown that the technique provides additional sensitivity to changing interfacial chemical states at interfaces. The X-ray standing wave provides a novel way to isolate the interfacial contribution to the detected signal that goes beyond conventional angle-resolved XPS. However, the questions that can be answered using this approach are somewhat limited by the choice of materials systems that have been employed.

Ideally, SW-XPS could be used to measure not simply changes in interfacial chemistry, but also to probe emergent interfacial electronic states near the Fermi level. There have also been some efforts to employ SW-XPS to generate ARPES information as a function of depth within the material, including studies of $SrTiO_3$-$GdTiO_3$[135] and $SrTiO_3$-$(La,Sr)MnO_3$[136] superlattices. These results are promising demonstrations of the technique, but provided limited insights into emergent interfacial physical phenomena due in large part to the relatively small modulation of the core level intensities (~20-30%). In most cases of SW-XPS studies to date, the challenge has been not the technique but the limitation of the samples themselves. We suggest that a somewhat different model could bear fruit.

Because the photon energies for most SW-XPS studies have been in the soft X-ray regime where it is possible to take advantage of elemental resonances to enhance reflectivity, the measurements are still limited to the top few nm of the sample. Thus, the underlying superlattice serves only to generate the X-ray standing wave that modulates the field strength across the top layer of the sample and does not have a meaningful effect on the actual data being measured. It therefore makes sense to choose an underlying superlattice solely from the perspective of generating the strongest standing wave intensity while not confusing the data interpretation from the relevant interface in other ways. In this sense, the superlattice serves as a novel substrate for the actual experimental interface. For example, a $SrTiO_3$-$BaTiO_3$ superlattice is relatively easy to synthesize and would provide strong standing wave effects across the Ba $M_5$ resonance (~780 eV). Predictions of superlattice Bragg peaks for various $BTO_n$-$STO_n$ superlattice structures using the X-ray Interactions with Matter website[137,138] provided by the Center for X-ray Optics at Lawrence Berkeley National Laboratory are shown in Figure 12(a). The electric field strength varies as the square root of reflectivity, meaning that values greater than 0.1 on the scale (dotted line) will produce modulations of XPS peak intensity of 30% or better. The superlattices could then be used as X-ray mirror substrates for any number of intriguing interfacial materials to produce much stronger standing-wave effects. A schematic of such a heterostructure is shown in Figure 12(b), with a $BTO_5$-$STO_5$ superlattice with an arbitrary $SrMO_3$-$SrM'O_3$ interface highlighted in yellow that will be strongly probed by the standing wave from the underlying mirror. Given this opportunity to shift the approach to standing-wave measurements, groups should consider the SW-XPS experiment when designing both their superlattice mirror layer and the interface of interest. Others have proposed additional approaches to enhance standing wave effects in these measurements[127]. Ideally, collaborations between synthesis groups and experts in SW-XPS early in the research process will produce samples that are particularly well suited to answer specific questions about the properties of an interface.

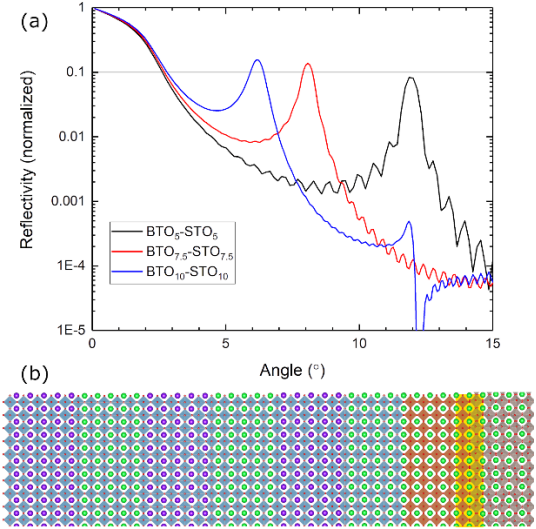

*Figure 12: (a) Modeled X-ray reflectivity for three n-unit cell BaTiO$_3$ (BTO$_n$)/n-unit cell SrTiO$_3$ (STO$_n$) superlattices on SrTiO$_3$ substrates at the Ba M$_5$ energy resonance; (b) Schematic of BTO$_5$-STO$_5$ superlattice mirror to probe an arbitrary SrBO$_3$-SrB'O$_3$ interface (highlighted in yellow).*

### C. Future Applications

In summary, synchrotron XPS techniques have generated significant new insights into buried oxide interfaces by taking advantage of the variable X-ray energies that are provided from a synchrotron source that cannot be replicated in a lab. Many of these approaches have only been pursued over the past decade, however, leaving significant opportunities to refine the practices to better extract novel physics from the experiments. For example, many beamlines have begun to integrate MBE or PLD growth systems *in situ* as has been done in a lab environment for many years[139,140]. We have shown in this review that surface exposure must be considered when interpreting results, even for HAXPES studies. It thus makes sense to perform SW-XPS and HAXPES studies using *in situ* growth capabilities at the beamline whenever possible. Alternatively, use of a vacuum suitcase to transport samples from a home synthesis laboratory to the user facility can enable synchrotron studies that replicate *in situ* conditions.

Another promising new opportunity has been enabled by the development of lab-based HAXPES instruments[141]. Rather than traditional Al or Mg sources, these commercially-available instruments employ a 3d transition metal or metalloid such as Cr or Ga to produce monochromatic Kα photons with energies greater than 5 keV. The recent development of a liquid-metal Ga source with a Rowland circle monochromator has produced fluxes that are within a factor of 200 of synchrotron sources, making it possible to perform measurements in reasonable times in a lab environment[141]. To date, we know of no laboratory that has an oxide or other complex materials film growth chamber attached *in situ* to a HAXPES instrument, but this would be an excellent resource for future film synthesis national or international user facilities.

## V. Summary

In conclusion, we have examined the state-of-the-art in oxide thin films and interfaces by focusing on the synergy that is possible when epitaxial film synthesis and X-ray photoelectron spectroscopy are integrated *in situ*. We have shown how XPS can be used to improve the synthesis process to observe film contamination, stoichiometry, surface termination, and surface chemistry. These day-to-day measurements can be extremely valuable to film growers as they optimize the growth process. We have also shown how more complex XPS experiments can be designed to measure interfacial intermixing, valence band offsets, and charge transfer. By examining several case studies, we showed the importance of designing thin film heterostructures with the subsequent XPS experiment in mind. Finally, we presented future directions for XPS-based studies of oxide interfaces that leverage advances in synchrotron and lab-based technologies to achieve high impact results. Bridging the gap between synthesis and characterization in these future studies will hopefully provide the oxide thin film community with a better understanding of the complex chemical and physical phenomena that drive emergent behavior in oxide heterostructures.

# VI.   Acknowledgements

We thank Shalinee Chikara for assistance with the $LaCoO_3$ deposition and XPS data acquisition shown in the paper. The authors gratefully acknowledge support from the Air Force Office of Scientific Research under award number FA9550-20-1-0034 and from the National Science Foundation Division of Materials Research under award number NSF-DMR-1809847. MB acknowledges fellowship support through the Alabama EPSCOR Graduate Research Scholars Program.